\documentclass[a4paper]{jpconf}
\usepackage{graphicx}
\DeclareRobustCommand{\ion}[2]{%
\relax\ifmmode
\ifx\testbx\f@series
{\mathbf{#1\,\mathsc{#2}}}\else
{\mathrm{#1\,\mathsc{#2}}}\fi
\else\textup{#1\,{\mdseries\textsc{#2}}}%
\fi}
\begin{document}
\title{Non-LTE spectral analyses of the lately discovered DB-gap white
  dwarfs from the SDSS}

\author{S. D. H\"ugelmeyer and S. Dreizler}

\address{Institut f\"ur Astrophysik, Georg-August-Universit\"at
  G\"ottingen, Friedrich-Hund-Platz~1, 37077~G\"ottingen, Germany}

\ead{shuegelm@astro.physik.uni-goettingen.de}

\begin{abstract} 
For a long time, no hydrogen-deficient white dwarfs have been known
that have effective temperature between $30~{\rm kK}$ and $< 45~{\rm
kK}$, i.\,e.\ exceeding those of DB white dwarfs and having lower ones
than DO white dwarfs. Therefore, this temperature range was long known
as the DB-gap. Only recently, the SDSS provided spectra of several
candidate DB-gap stars. First analyses based on model spectra
calculated under the assumption of local thermodynamic equilibrium
(LTE) confirmed that these stars had $30~{\rm kK} < T_{\rm eff} <
45~{\rm kK}$ (Eisenstein et al. 2006). It has been shown for DO white
dwarfs that the relaxation of LTE is necessary to account for non
local effects in the atmosphere caused by the intense radiation
field. Therefore, we calculated a non-LTE model grid and re-analysed
the aforementioned set of SDSS spectra. Our results confirm the
existence of DB-gap white dwarfs.
\end{abstract}

\section{Introduction}
The cooling sequence of hydrogen-deficient white dwarfs (WDs) is
populated by different subtypes of this species. At $T_{\rm eff} >
45~{\rm kK}$, stars are called DO WDs. These objects display mainly
\ion{He}{ii} at high and a combination of \ion{He}{ii} and \ion{He}{i}
lines at lower temperatures. Stars with spectra showing only
\ion{He}{i} lines are called DB WDs. Prior to the data releases of the
Sloan Digital Sky Survey (SDSS), the hottest DB white dwarf (WD) known
was PG~0112+104 with $\sim 30\,000~{\rm K}$. The coolest DO WD prior
to the SDSS on the other hand was PG~1133+489 with $47\,500~{\rm K}$
(Wesemael et al. 1985). These two stars constituted the cool and hot
end of the so-called DB-gap (Liebert et al. 1986), a temperature
region with a deficiency in helium-rich white dwarfs. The SDSS survey
was rich in new white dwarfs and brought forth a large number of DB
and DO white dwarfs. In the paper of Eisenstein et al. (2006), the
authors report a considerable number of H-deficient WDs with
temperatures that put them in the DB-gap. To derive WD temperatures,
they used model atmospheres assuming local thermodynamic equilibrium
(LTE). Dreizler \& Werner (1996) showed that even for stars with
$T_{\rm eff} = 50~{\rm kK}$ non-local effects are still
important. Therefore we re-analysed the sample of Eisenstein et
al. (2006) with our non-LTE model spectra.

\section{Models and fitting}
For our non-LTE spectral analysis, we have calculated a grid of model
atmospheres with the following parameters:

\vspace{.5cm}

\begin{tabular}{l c l c l}
  $T_{\rm eff}:$ &&  $27\,500 - 50\,000~{\rm K}$ && ${\rm steps\ of}\
  2\,500~{\rm K}$\\
  $\log{g}:$ && $7.60 - 8.80~{\rm (cgs)}$ && ${\rm steps\ of}\ 0.20~{\rm
    dex}$\\
  $X({\rm He}):$ && $99.0\ \&\ 99.9$ && 
\end{tabular}

\vspace{.5cm}

Our fitting method applied is the same as described in
H{\"u}gelmeyer et al. (2006). It is similar to the one used by
Koester in Eisenstein et al. (2006). We first normalized
the SDSS spectrum by determining the continuum points of the
observations by the continuum of the normalized model spectrum and
then fitted a third order polynomial to the double logarithmic data.
We then applied a $\chi^2$-fit to the normalized models in certain
wavelength regions, i.\,e.\ $3900$~\AA\,--\,$5100$~\AA,
$5800$~\AA\,--\,$6000$~\AA, and $6500$~\AA\,--\,$6800$~\AA.

\section{Results \& discussion}
Our best fit models are shown in Figure~2 and the photospheric
values are presented in Table~1. The comparison of our NLTE model
fits to the temperature values obtained with LTE Koester models is
shown in Figure~1. Our derived temperatures are in better agreement
with the fitting method of Koester. This is not very surprising
since the procedures are very similar. The Eisenstein
\texttt{autofit} method fits photometric data additionally to the
spectroscopy.

The left panel plot in Figure~1 shows that our fitting method,
especially in the high temperature regime, produces higher $T_{\rm
eff}$ values than the Eisenstein method. However, all but two fits --
one of which is for an object with a significant amount of hydrogen
and therefore cannot be reliably fit with our He-rich models-- are
within the 15\% deviation range. The Koester fits are in even better
agreement with our model fitting. Direct comparison of our non-LTE to
the LTE Koester models show some obvious deviations in the continuum
flux which increase towards lower temperatures. Currently, we do not
have an explanation for these discrepancies.

Overall, we can confirm the results of Eisenstein et
al. (2006). Especially the nice agreement of our fits to those of
Koester at effective temperatures $>45~{\rm K}$ suggests that LTE is a
valid assumption for the analysis of DB-gap white dwarfs.
  
\begin{figure}
    \includegraphics[width=0.48\textwidth]{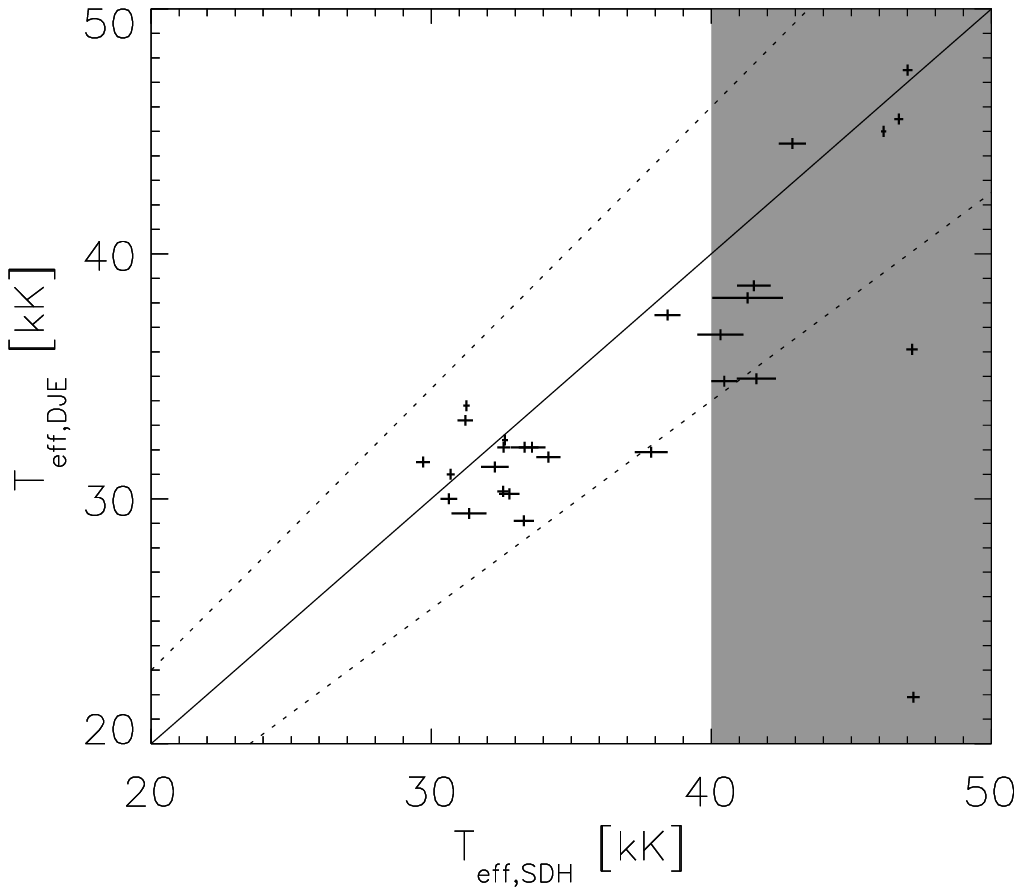}
    \includegraphics[width=0.48\textwidth]{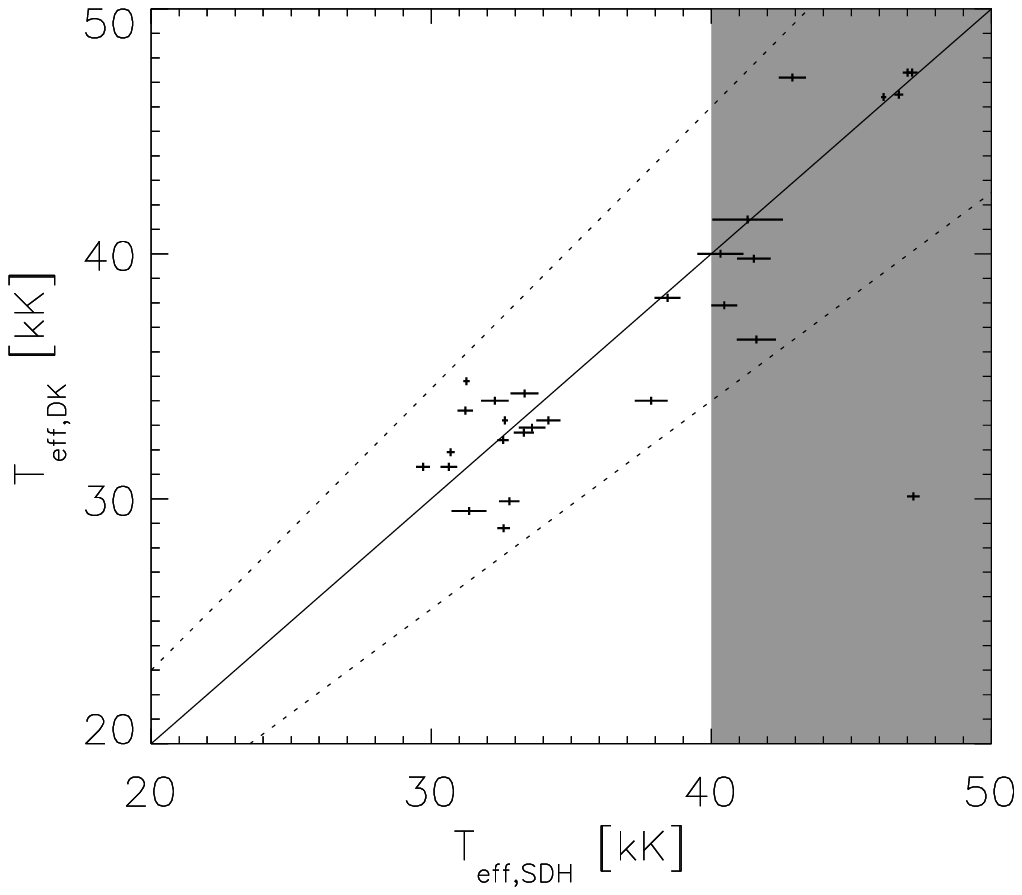} 
    \caption{Comparison of LTE effective temperatures from the
    Eisenstein (left panel) and Koester (right panel) fits to our NLTE fits
    for all analysed WDs. The Eisenstein fits show systematically
    lower temperatures than the Koester fits. The latter ones agree
    better with our fitting results. The shaded area is the
    temperature range where we expect deviations from LTE, i.~e.\
    differences in fitted temperatures. The dotted lines are 15\%
    deviations from equality (solid lines).}
\end{figure}

\begin{table}[!h]
  \centering
  \caption{Best-fit model values for the analyzed SDSS
    spectra. SDSS~J153852.35-012133.8 clearly shows hydrogen Balmer lines
    and cannot be fit with our model grid. The real effective temperature
    is lower than the value given in this table. All errors are
    $1$-$\sigma$ statistical errors.} 
  \smallskip
  \begin{tabular}{l c l@{\,$\pm$\,}r c l@{\,$\pm$\,}r c c c c}
    \hline\hline\\
    \multicolumn{1}{c}{Name} & & \multicolumn{2}{c}{$T_{\rm eff}~{\rm
        [K]}$} & & \multicolumn{2}{c}{$\log{g}$ (cgs)} & & $X({\rm
      He})$ & & $\chi^2$ \\[3pt] \hline
    SDSS J001529.75+010521.4& &40325&821& &7.88&0.07& & 99.9 & & 1.1601\\
    SDSS J040854.60$-$043354.7& &41607&697& &7.76&0.04& & 99.0 & &1.2157\\
    SDSS J074538.17+312205.4& &42890&483& &7.72&0.06& & 99.9 & & 1.2050\\
    SDSS J081115.09+270621.8& &47007&168& &7.91&0.03& & 99.0 & &1.1289\\
    SDSS J081546.08+244603.3& &46692&157& &7.92&0.04& & 99.0 & &1.1678\\
    SDSS J084823.53+033216.7& &33327&495& &7.75&0.07& & 99.0 & &1.1150\\
    SDSS J084916.18+013721.3& &31347&618& &8.03&0.04& & 99.9 & & 1.1431\\
    SDSS J090232.18+071930.0& &32787&360& &7.96&0.03& & 99.9 & & 1.1815\\
    SDSS J090456.13+525029.9& &38435&461& &7.60&0.01& & 99.0 & &1.2014\\
    SDSS J092544.41+414803.2& &41517&596& &7.83&0.10& & 99.9 & & 1.1468\\
    SDSS J093041.80+011508.4& &37850&585& &7.60&0.01& & 99.9 & & 1.1464\\
    SDSS J093759.52+091653.3& &29705&247& &8.23&0.01& & 99.9 & & 1.2879\\
    SDSS J095256.69+015407.7& &31257&112& &8.23&0.01& & 99.9 & & 1.2535\\
    SDSS J113609.59+484318.9& &46152&90& &8.17&0.01& & 99.9 & & 1.4391\\
    SDSS J123750.47+085526.0& &32270&495& &8.24&0.03& & 99.9 & & 1.1364\\
    SDSS J134524.92$-$023714.2& &41292&1260& &7.96&0.03& & 99.9 & & 1.2249\\
    SDSS J140159.09+022126.7& &40460&461& &7.99&0.03& & 99.9 & & 1.1346\\
    SDSS J141258.17+045602.2& &30695&135& &8.19&0.01& & 99.9 & & 1.3531\\
    SDSS J141349.46+571716.4& &30627&292& &8.17&0.02& & 99.9 & & 1.1883\\
    SDSS J143227.25+363215.2& &33305&360& &7.91&0.04& & 99.9 & & 1.1576\\
    SDSS J153852.35$-$012133.8& &47210&225& &7.60&0.01& & 99.0 & &1.1996\\
    SDSS J154201.50+502532.1& &32630&101& &7.96&0.01& & 99.9 & & 1.4335\\
    SDSS J164703.44+245129.1& &34182&427& &7.60&0.01& & 99.9 & & 1.1683\\
    SDSS J211149.60$-$053938.4& &47165&202& &7.74&0.05& & 99.0 & &1.1400\\
    SDSS J212403.12+114230.2& &32562&202& &7.99&0.02& & 99.9 & & 1.2352\\
    SDSS J215514.44$-$075833.8& &32585&225& &8.05&0.02& & 99.9 & & 1.1418\\
    SDSS J222833.82+141037.0& &31212&270& &8.32&0.03& & 99.9 & & 1.1198\\
    SDSS J234709.29+001858.0& &33597&472& &7.96&0.05& & 99.9 & & 1.2369\\
    \hline
  \end{tabular}
\end{table}

\begin{figure}[!h]
  \centering
  \includegraphics[bb=-13 -13 510 161,clip,width=0.31\textwidth]{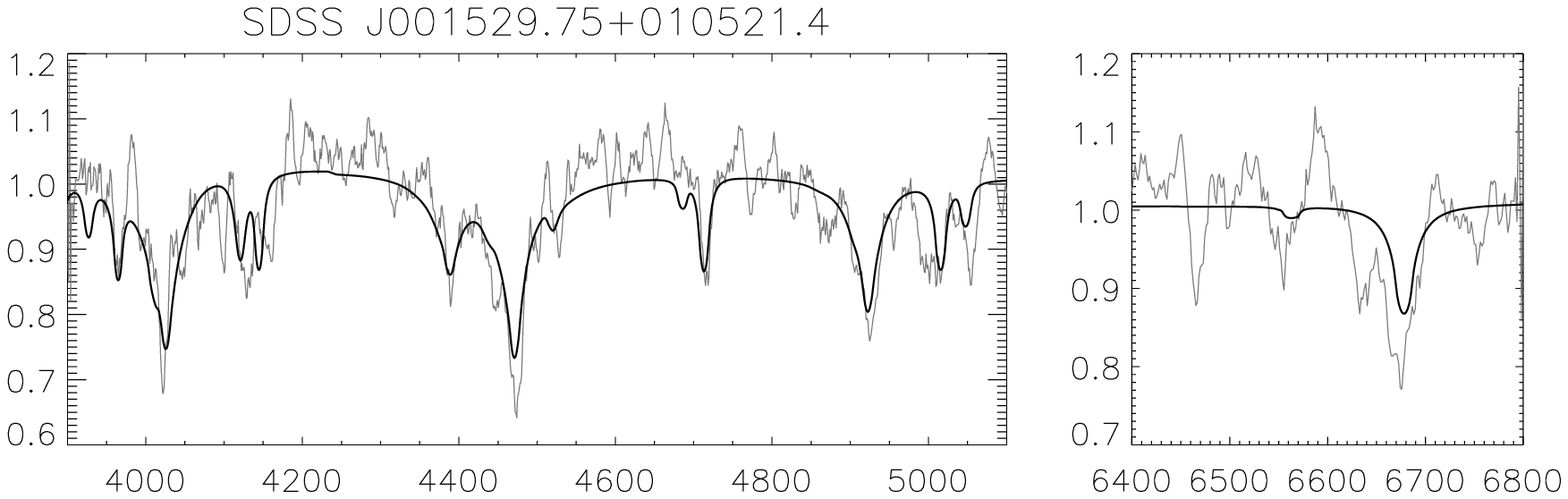}
  \includegraphics[bb=-13 -13 510 161,clip,width=0.31\textwidth]{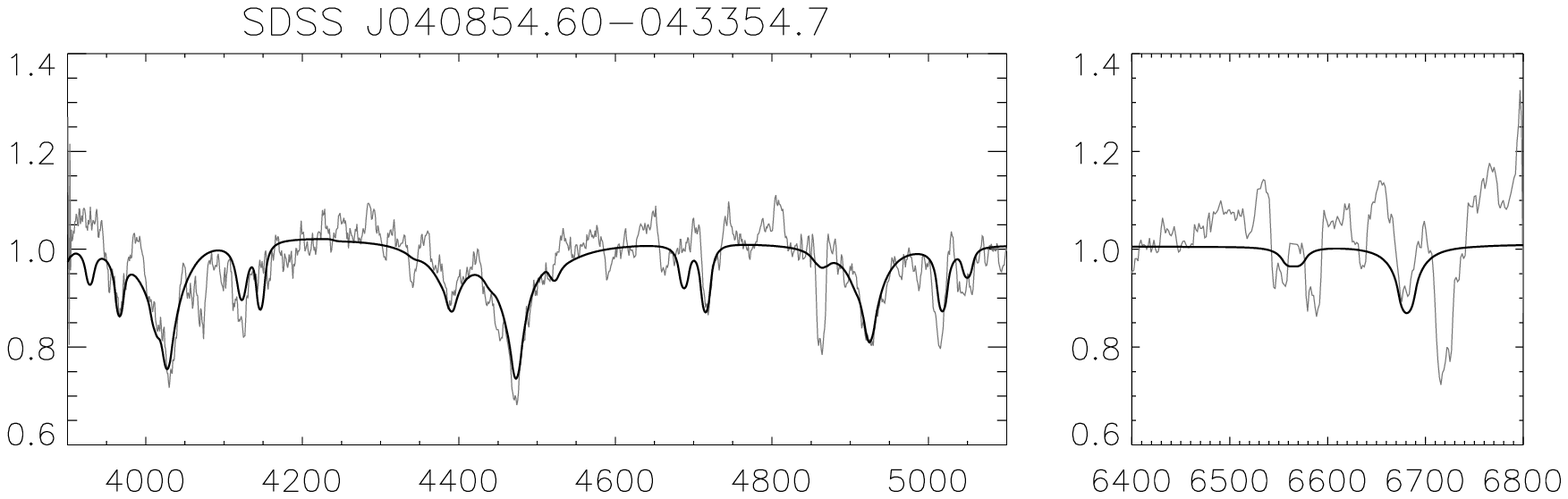}
  \includegraphics[bb=-13 -13 510 161,clip,width=0.31\textwidth]{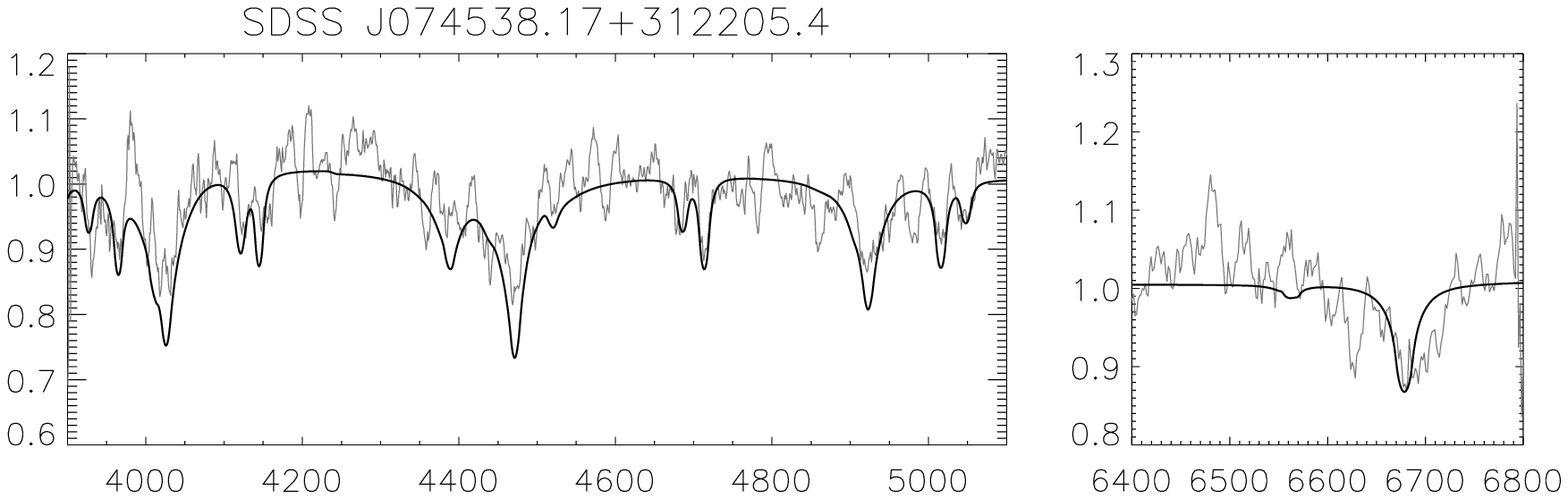}
  \includegraphics[bb=-13 -13 510 161,clip,width=0.31\textwidth]{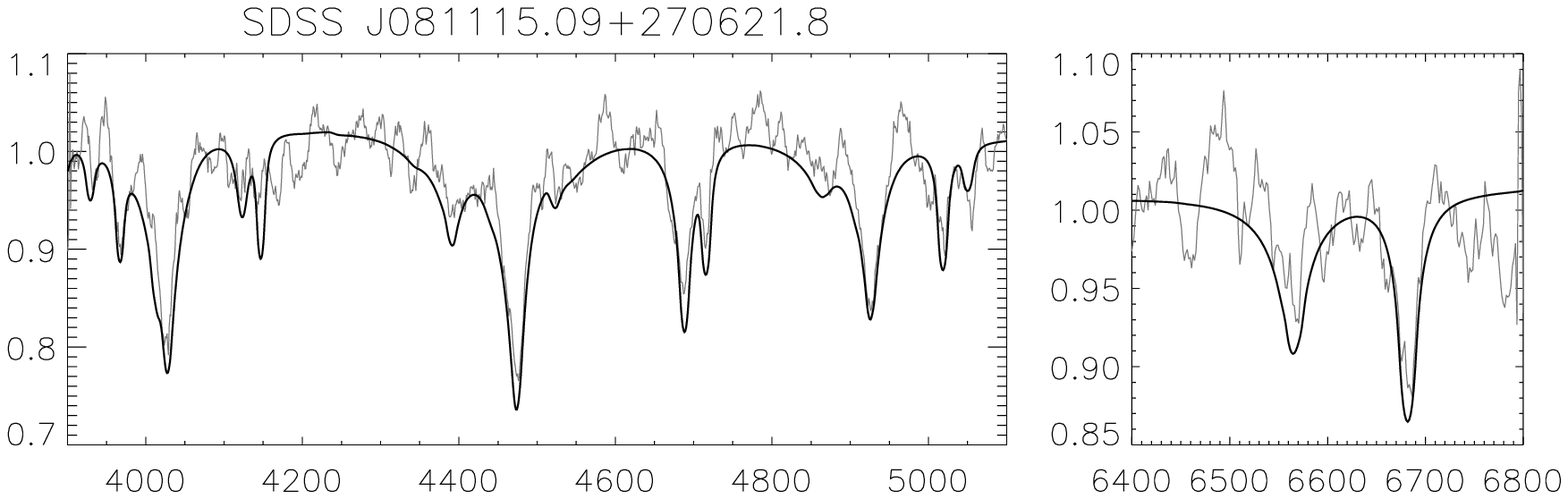}
  \includegraphics[bb=-13 -13 510 161,clip,width=0.31\textwidth]{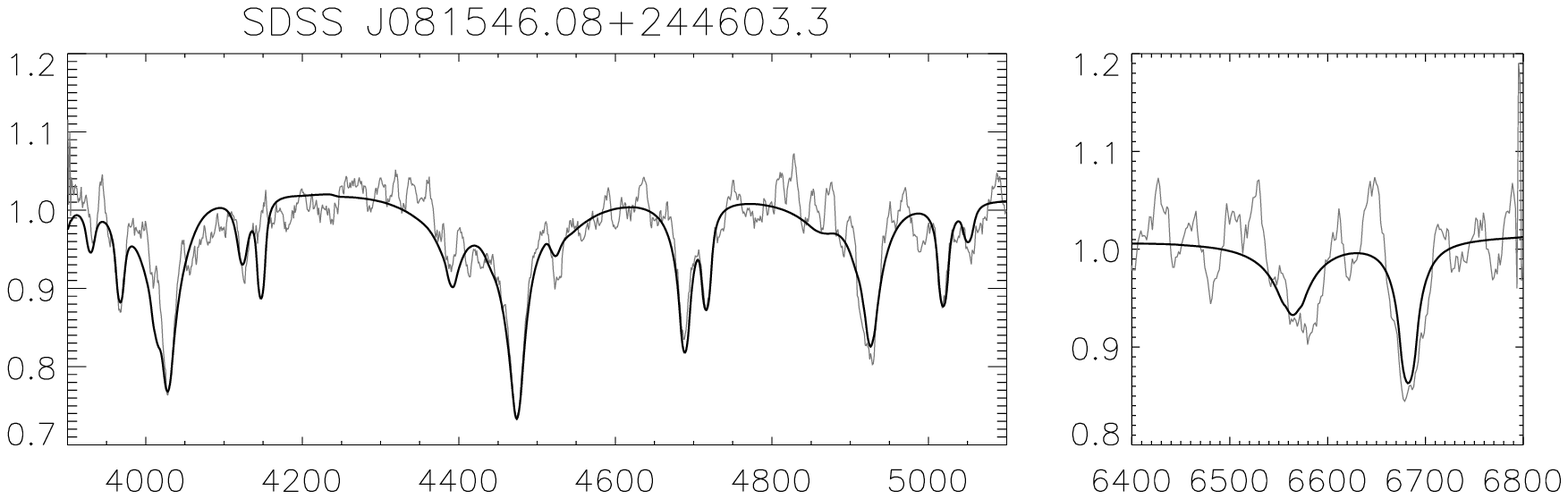}
  \includegraphics[bb=-13 -13 510 161,clip,width=0.31\textwidth]{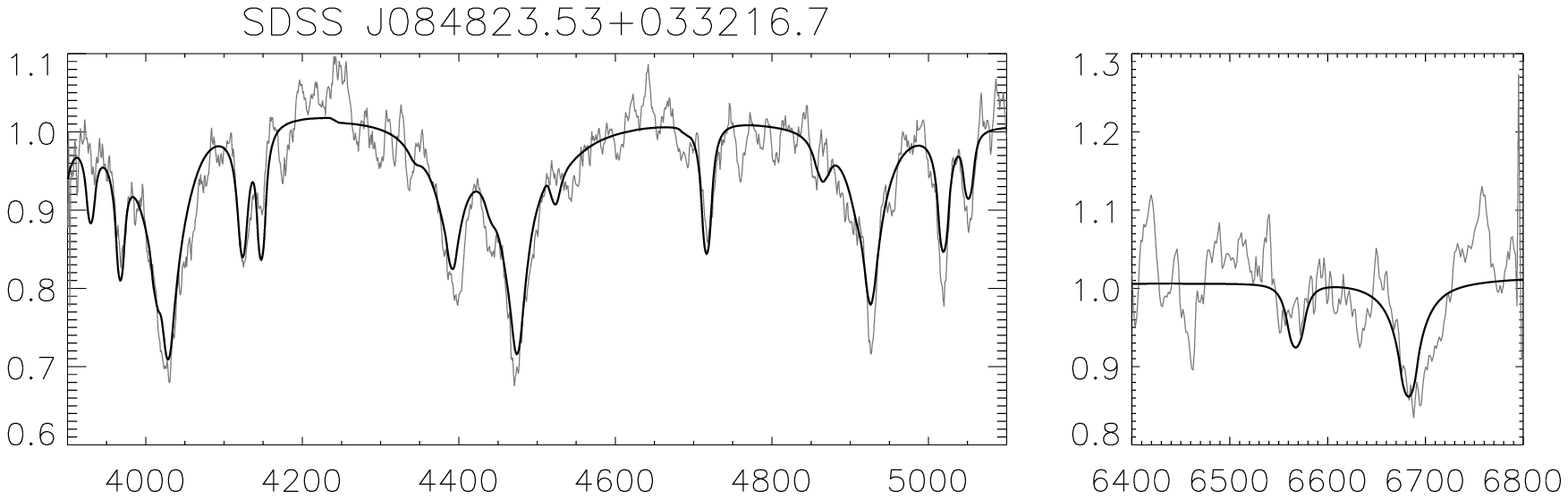}
  \includegraphics[bb=-13 -13 510 161,clip,width=0.31\textwidth]{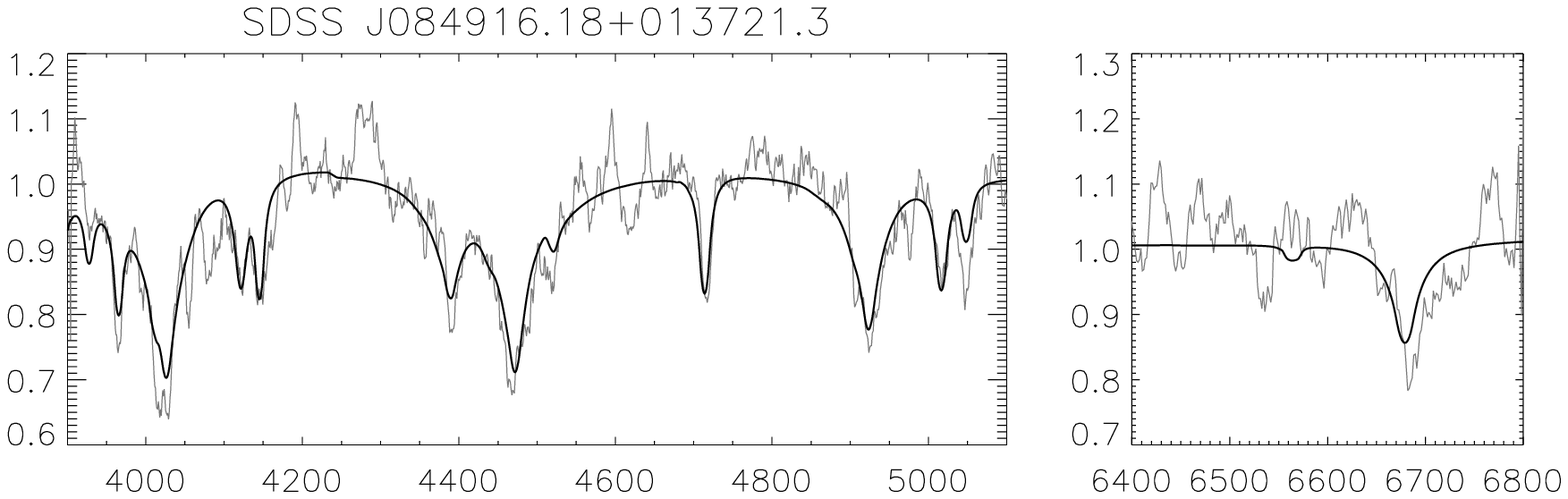}
  \includegraphics[bb=-13 -13 510 161,clip,width=0.31\textwidth]{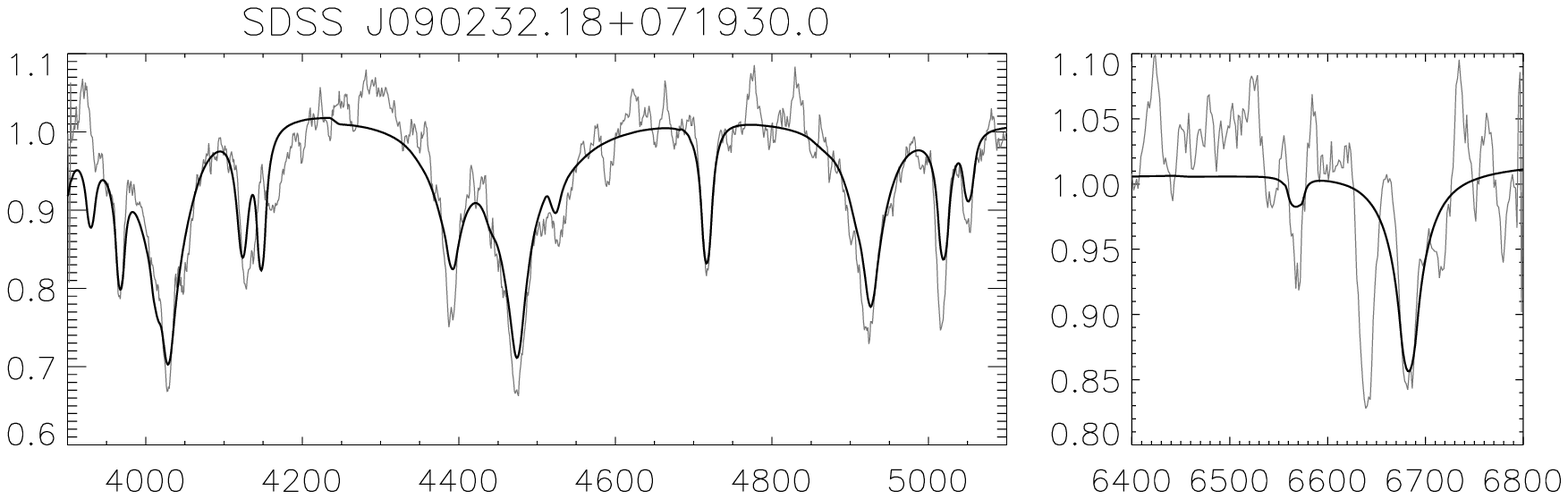}
  \includegraphics[bb=-13 -13 510 161,clip,width=0.31\textwidth]{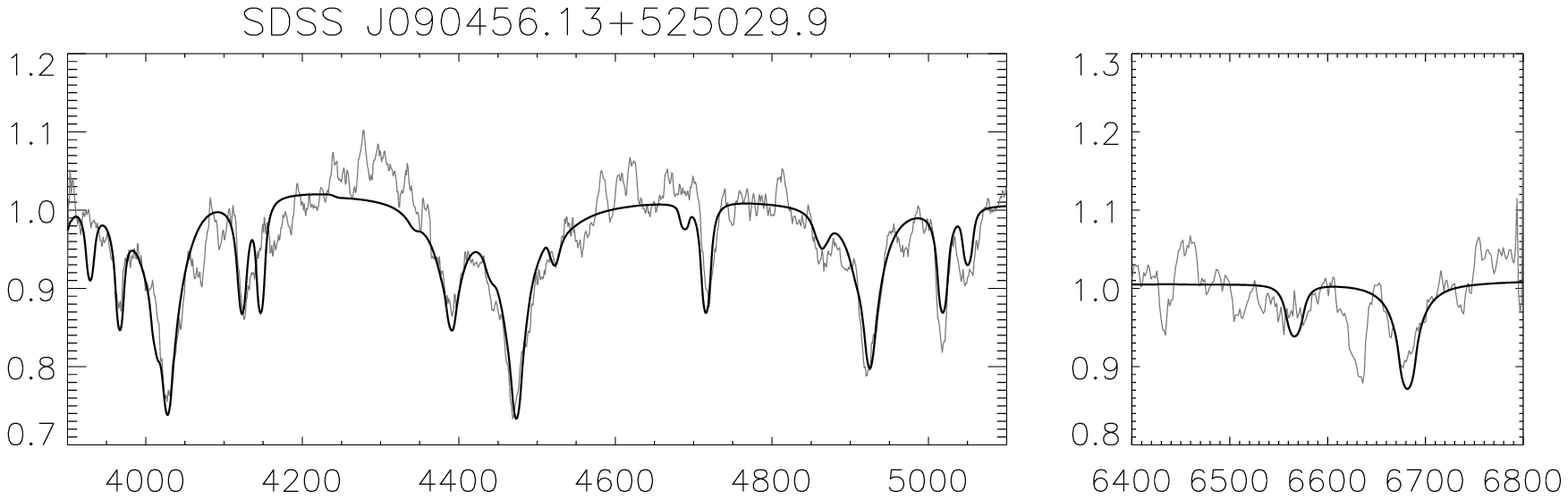}
  \includegraphics[bb=-13 -13 510 161,clip,width=0.31\textwidth]{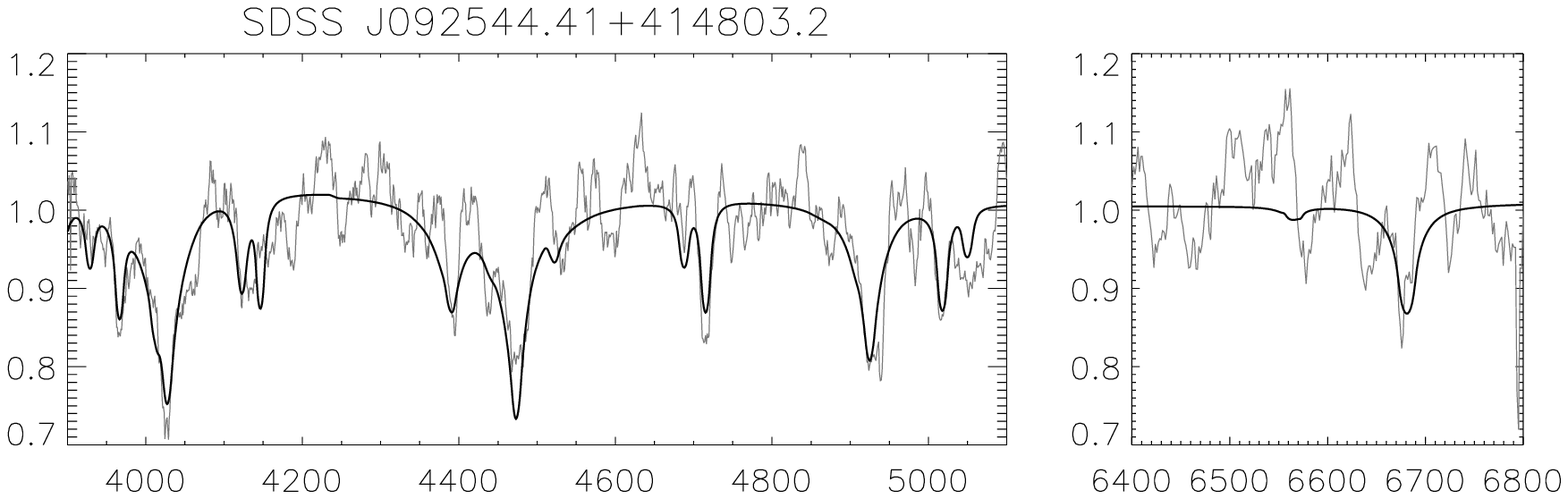}
  \includegraphics[bb=-13 -13 510 161,clip,width=0.31\textwidth]{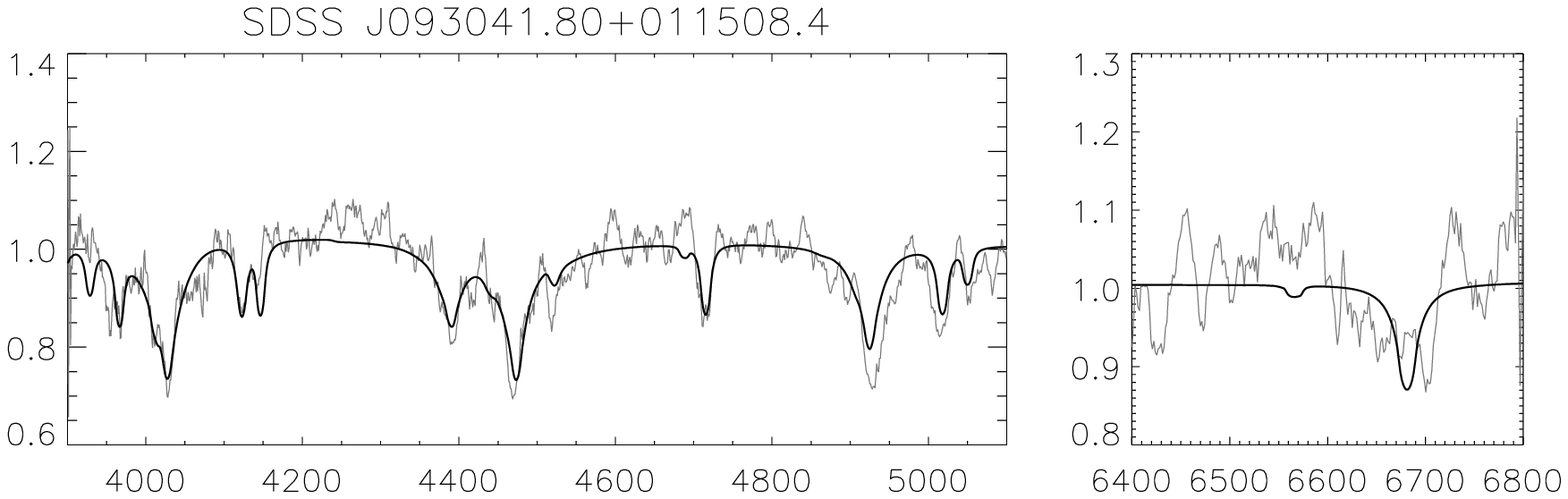}
  \includegraphics[bb=-13 -13 510 161,clip,width=0.31\textwidth]{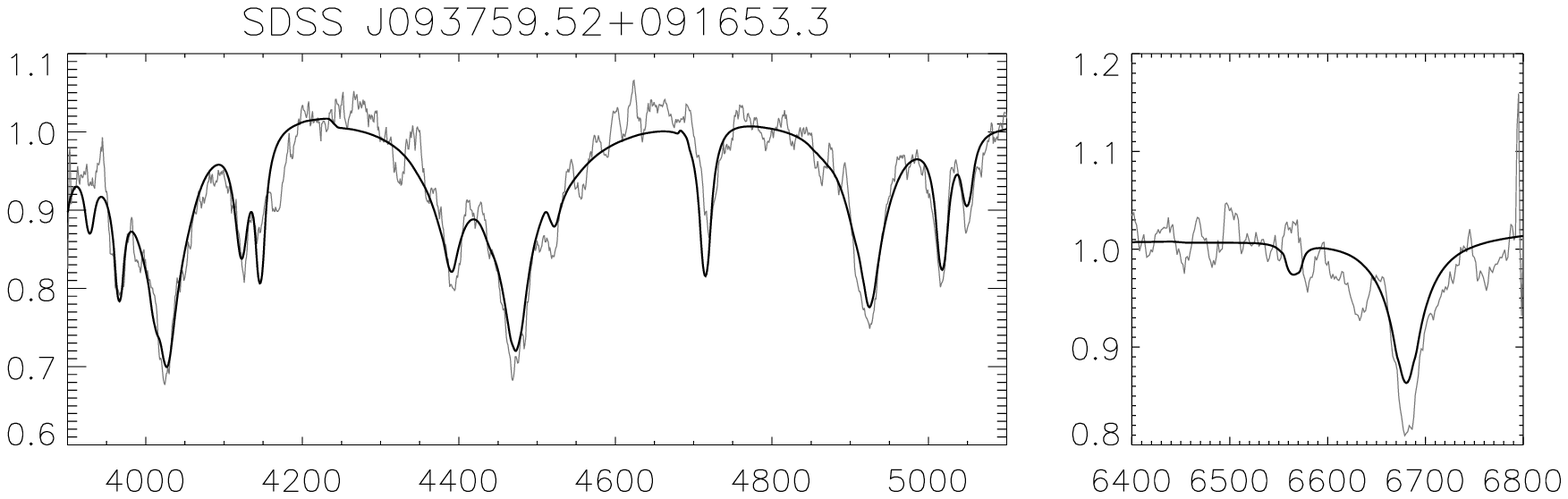}
  \includegraphics[bb=-13 -13 510 161,clip,width=0.31\textwidth]{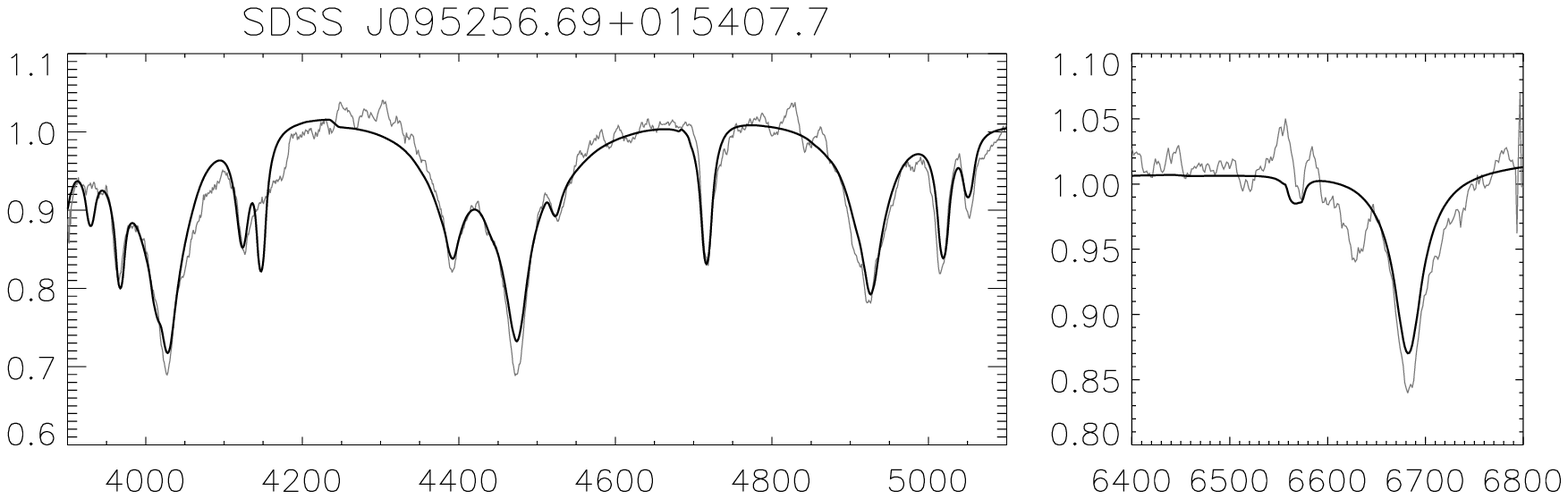}
  \includegraphics[bb=-13 -13 510 161,clip,width=0.31\textwidth]{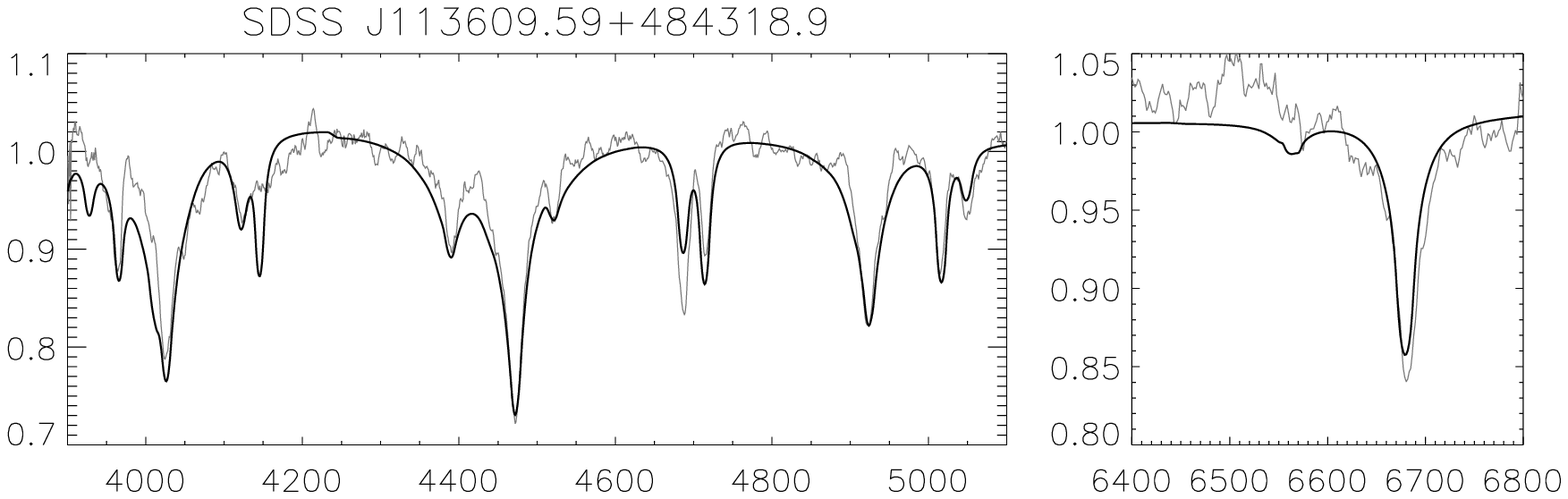}
  \includegraphics[bb=-13 -13 510 161,clip,width=0.31\textwidth]{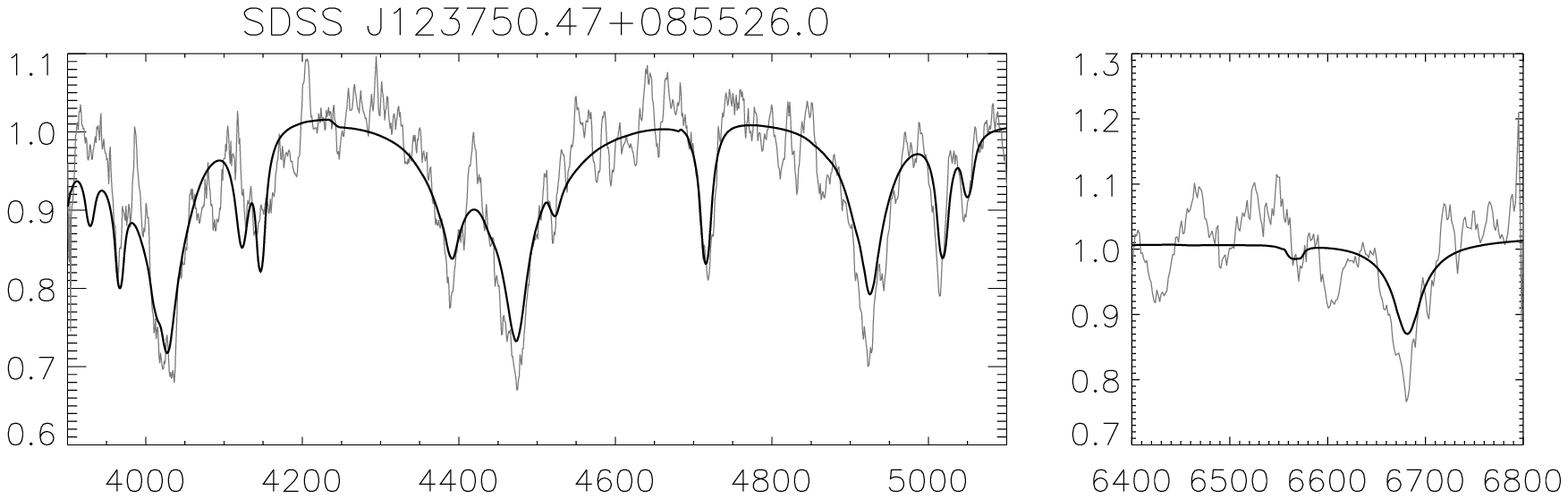}
  \includegraphics[bb=-13 -13 510 161,clip,width=0.31\textwidth]{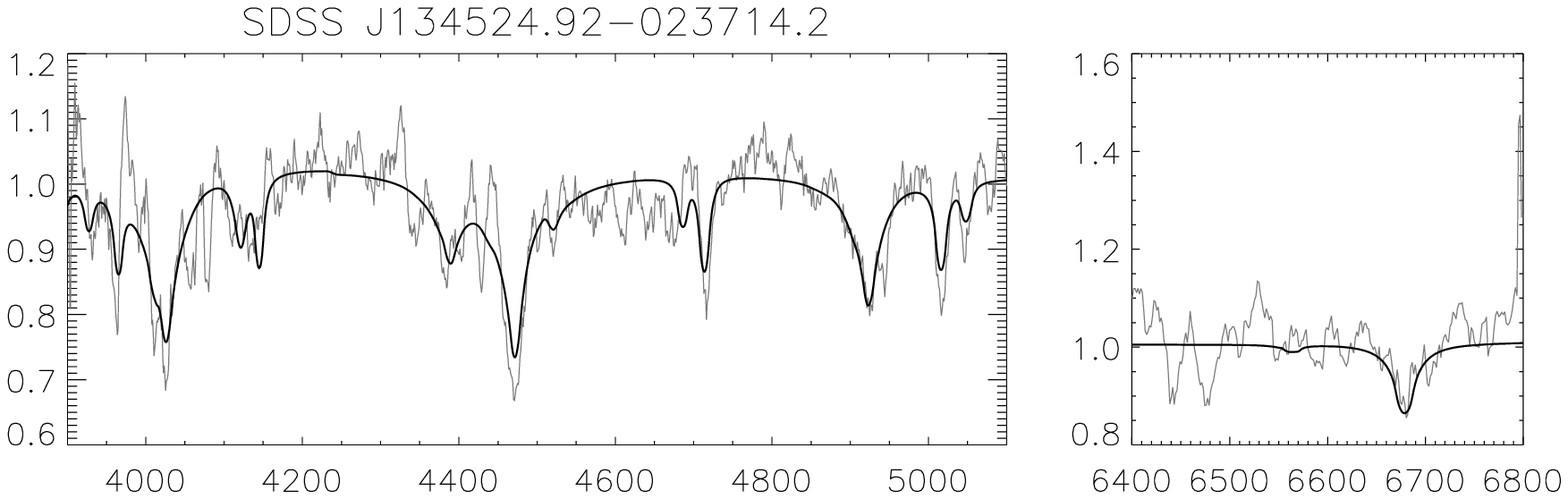}
  \includegraphics[bb=-13 -13 510 161,clip,width=0.31\textwidth]{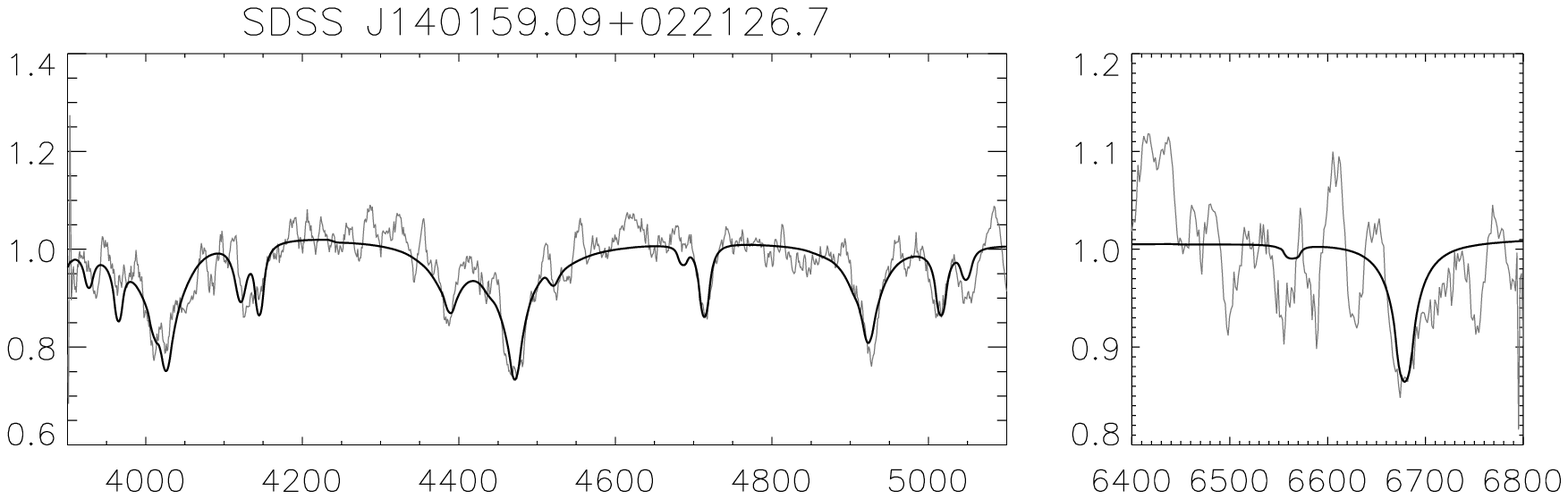}
  \includegraphics[bb=-13 -13 510 161,clip,width=0.31\textwidth]{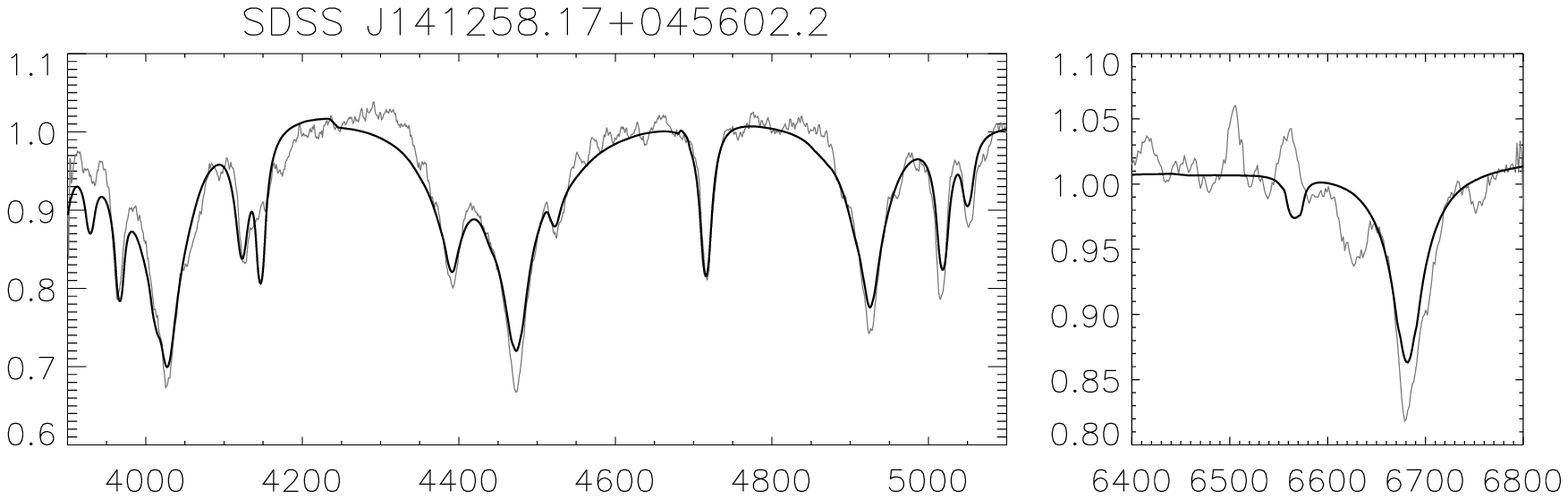}
  \includegraphics[bb=-13 -13 510 161,clip,width=0.31\textwidth]{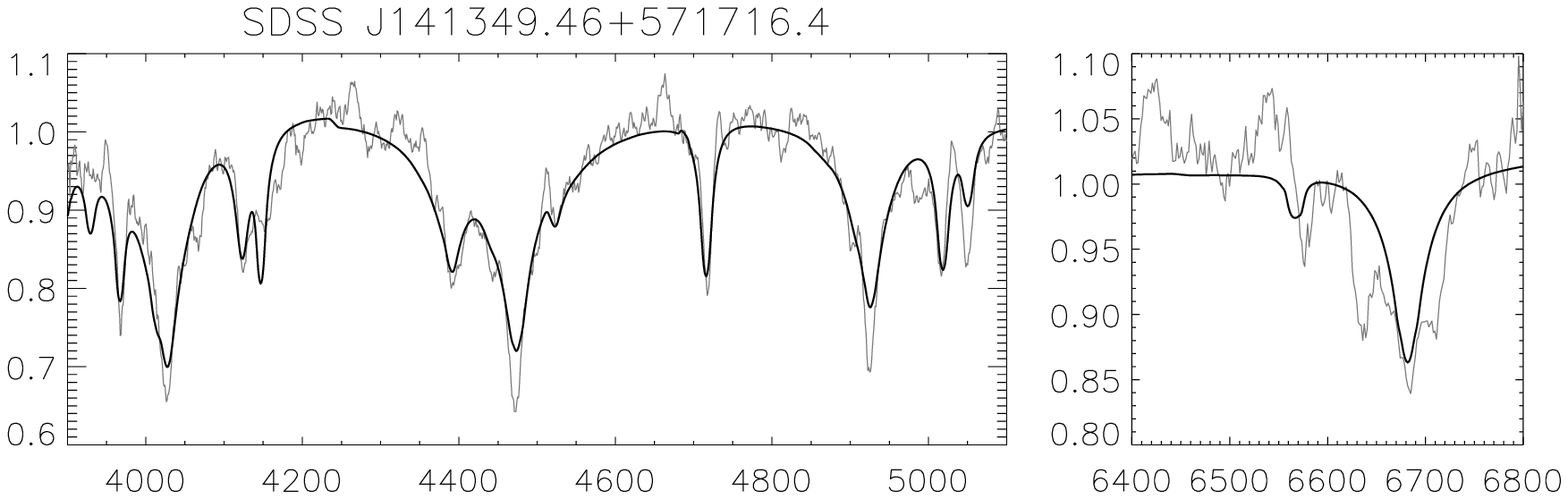}
  \includegraphics[bb=-13 -13 510 161,clip,width=0.31\textwidth]{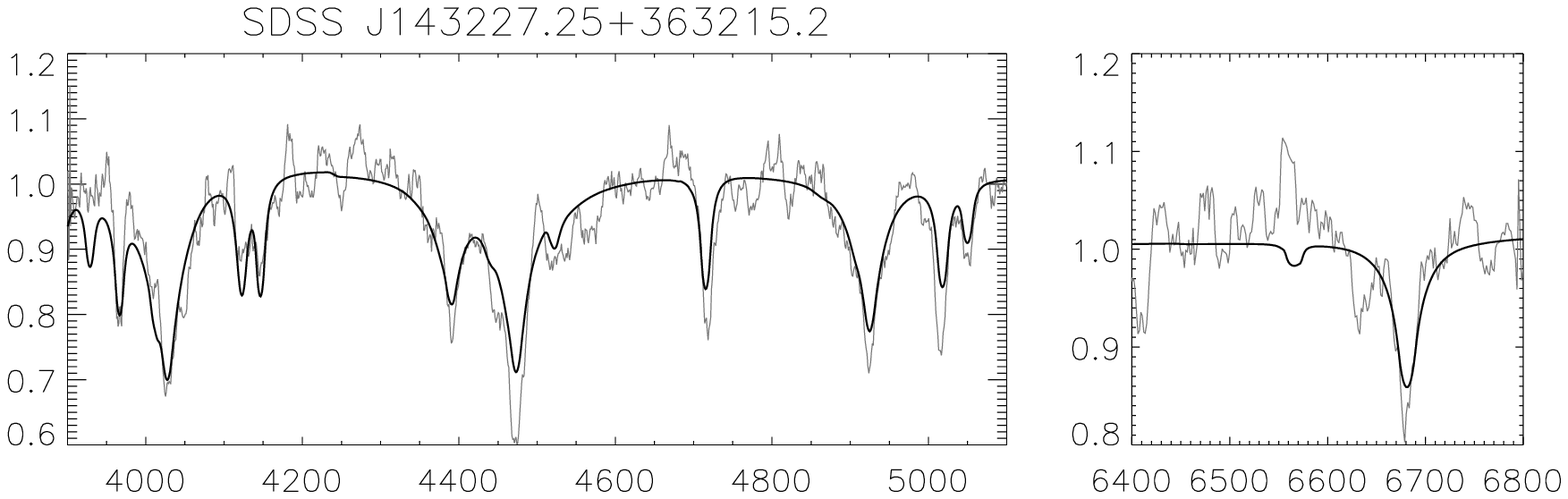}
  \includegraphics[bb=-13 -13 510 161,clip,width=0.31\textwidth]{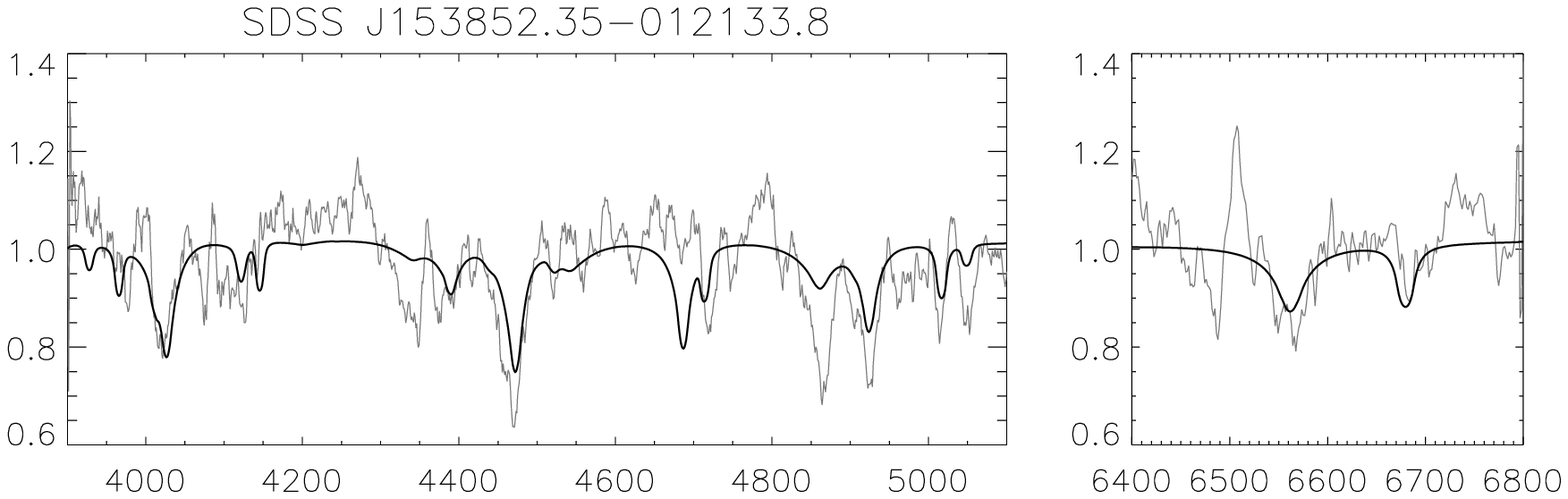}
  \includegraphics[bb=-13 -13 510 161,clip,width=0.31\textwidth]{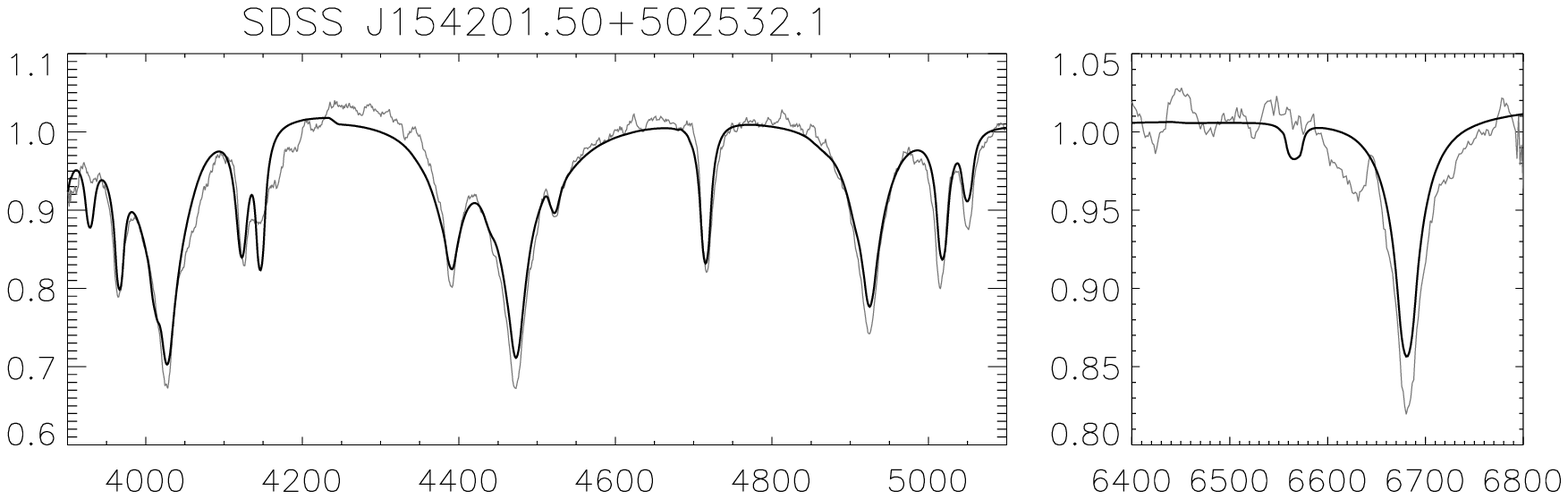}
  \includegraphics[bb=-13 -13 510 161,clip,width=0.31\textwidth]{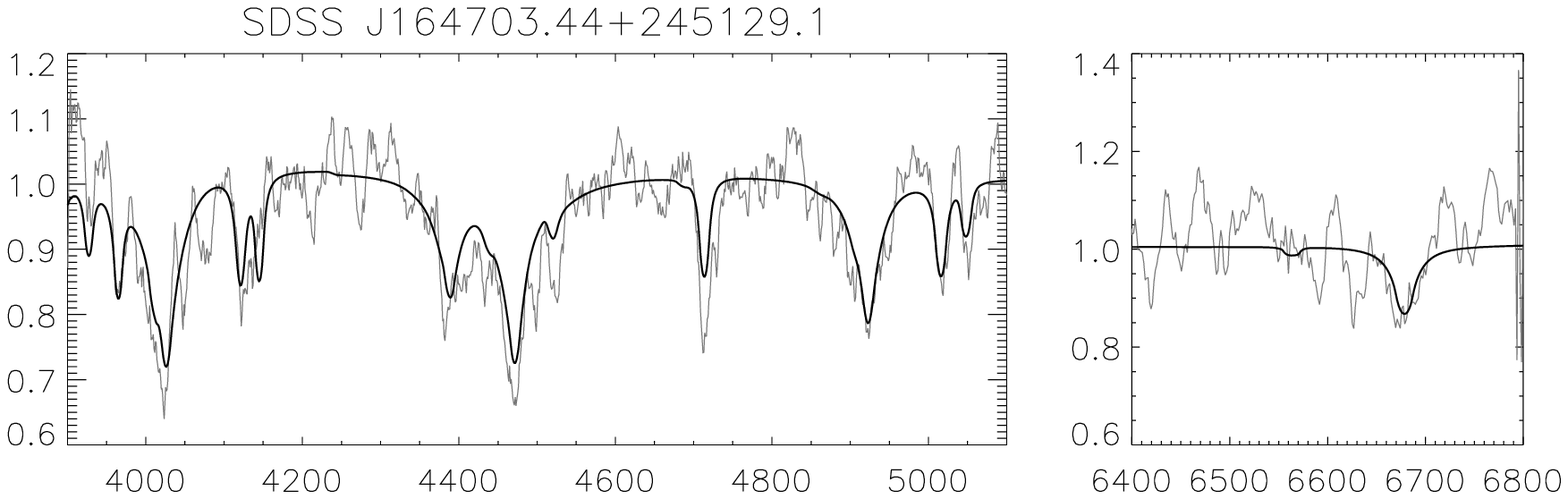}
  \includegraphics[bb=-13 -13 510 161,clip,width=0.31\textwidth]{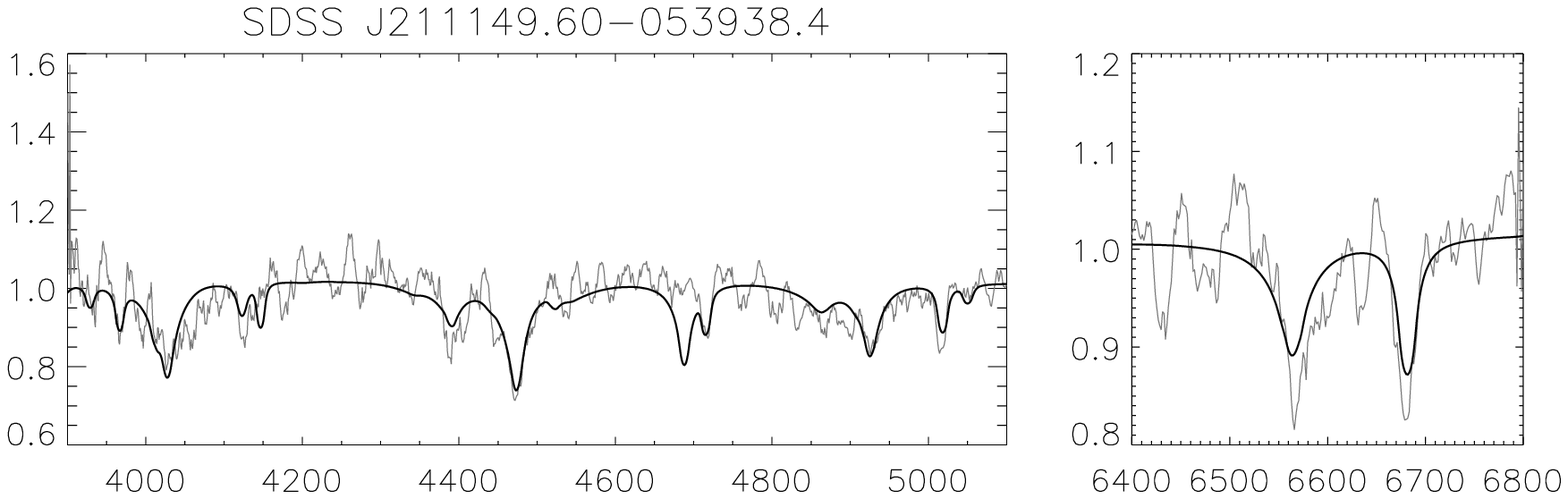}
  \includegraphics[bb=-13 -13 510 161,clip,width=0.31\textwidth]{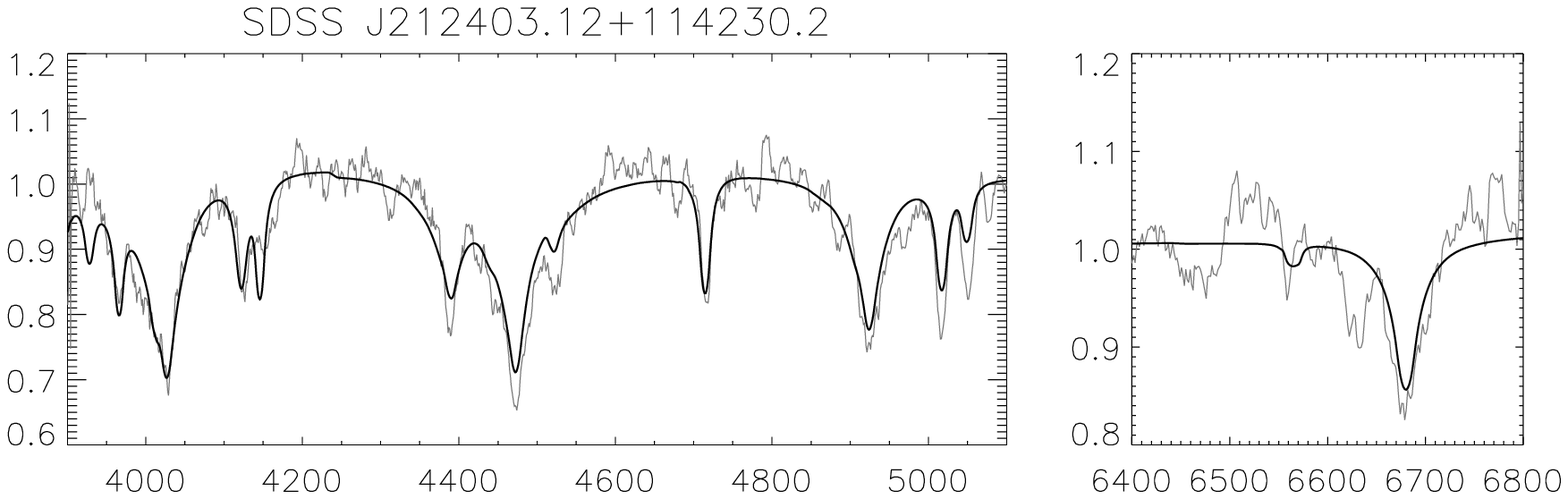}
  \includegraphics[bb=-13 -13 510 161,clip,width=0.31\textwidth]{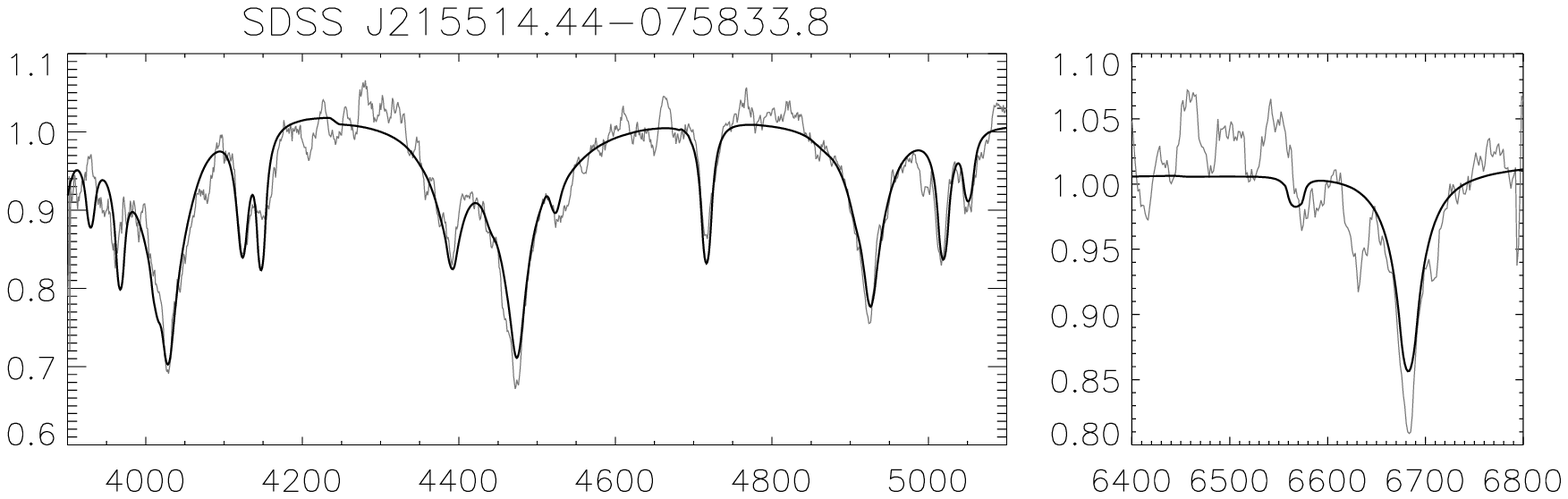}
  \includegraphics[bb=-13 -13 510 161,clip,width=0.31\textwidth]{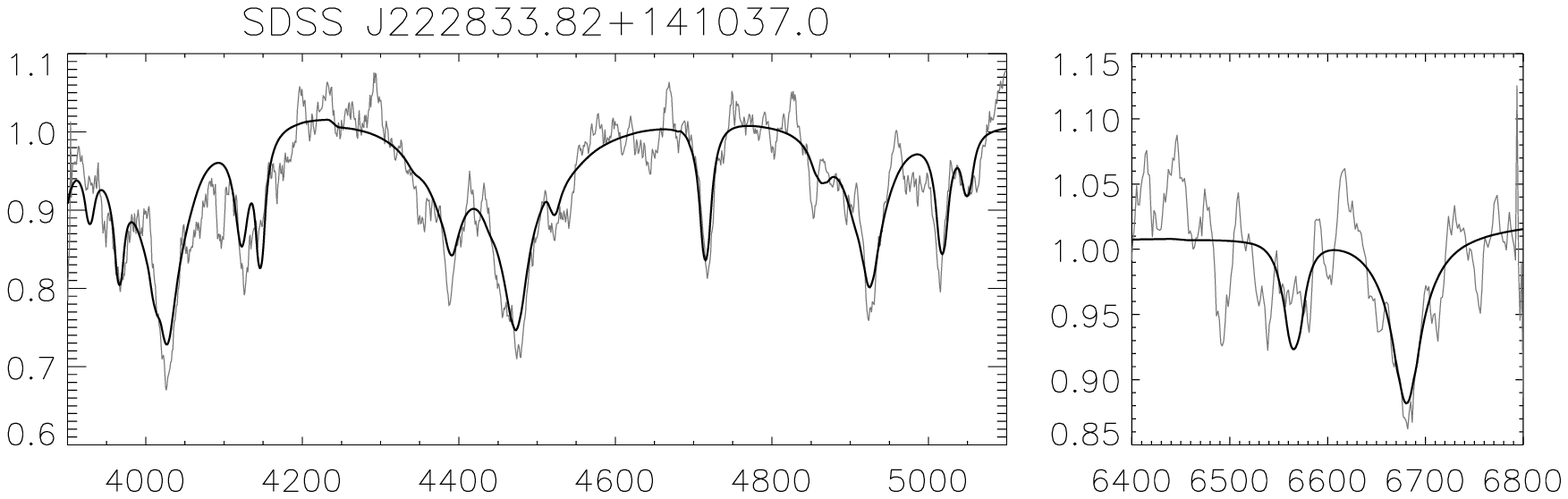}
  \includegraphics[bb=-13 -13 510 161,clip,width=0.31\textwidth]{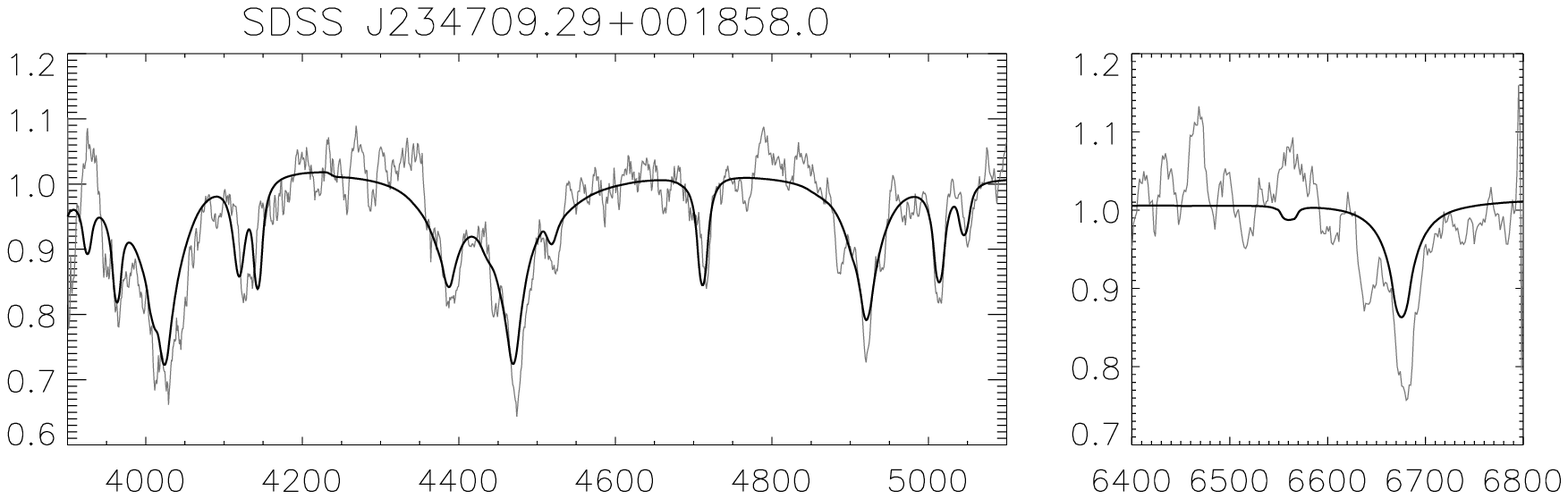}
  \caption{SDSS spectra (grey lines) and best-fit models (black
    lines). SDSS~J153852.35-012133.8 clearly shows hydrogen Balmer lines
    and cannot be fit with our model grid. Instead the fitting method
    tries to compensate the missing hydrogen in the model by increasing
    effective temperature to fill H$\alpha$ with He${\rm II}$. For
    clarity the SDSS spectra are smooth for these plots.}      
  \label{fig:spec}
\end{figure}

\section*{Acknowledgments} SDH is supported by a
  scholarship of the DFG Graduiertenkolleg 1351 ``Extrasolar Planets
  and their Host Stars''

\section*{References}

\begin{thereferences}
\item {Dreizler}, S. \& {Werner}, K. 1996 {\sl A\&A} {\bf 314} 217
\item {Eisenstein}, D.~J., {Liebert}, J., {Koester}, D., et
  al. 2006 {\sl AJ} {\bf 132} 676
\item {H{\"u}gelmeyer}, S.~D., {Dreizler}, S., {Homeier}, D., et
  al. 2006 {\sl A\&A} {\bf 454} 617
\item {Liebert}, J., {Wesemael}, F., {Hansen}, C.~J., et al. 1986
  {\sl ApJ} {\bf 309} 241
\item {Wesemael}, F., {Green}, R.~F., \& {Liebert}, J. 1985 {\sl
    ApJS} {\bf 58} 379
\end{thereferences}


\end{document}